\documentclass{appolb}

\usepackage{graphicx}
\usepackage{bm} 
\usepackage{subfigure}

\begin{document}

\title{Initial conditions for hydrodynamics: implications for phenomenology%
\thanks{Presented by WB at the IV Workshop on Particle Correlations and Femtoscopy, Cracow, 11-14 September 2008}%
}
\author{Wojciech Broniowski$^{a,b}$, Wojciech Florkowski$^{a,b}$,\\ Miko\l{}aj Chojnacki$^a$, and Adam Kisiel$^{c,d}$
\address{$^a$The H. Niewodnicza\'nski Institute of Nuclear Physics,\\ Polish Academy of Sciences, PL-31342 Krak\'ow, Poland\\
         $^b$Institute of Physics, Jan Kochanowski University, PL-25406~Kielce, Poland\\
         $^c$Faculty of Physics, Warsaw University of Technology, PL-00661 Warsaw, Poland\\
         $^d$Department of Physics, Ohio State University, Columbus, OH 43210, USA  }
}

\date{29 December 2008}

\maketitle

\pagestyle{empty}

\begin{abstract}
It is shown how the initial azimuthally asymmetric flow develops from the free-streaming + sudden equilibration approximation to the early 
dynamics in relativistic heavy-ion collisions. Consequences for the hydrodynamics and physical results are discussed. 
\end{abstract}


\bigskip  

We present a description of the early-stage dynamics in relativistic heavy ion collisions, where the free-streaming (FS) 
of partons is followed by a sudden equilibration (SE) to a thermalized phase, providing initial conditions for the subsequent 
hydrodynamic evolution. This FS+SE approximation has been proposed by Kolb, Sollfrank, 
and Heinz \cite{Kolb:2000sd}. It has been further discussed in the context of the 
isotropization problem by Jas and Mr\'owczy\'nski \cite{Jas:2007rw}, as well as used to analyze the 
early development of collective flow by Sinyukov, Gyulassy, Karpenko, and Nazarenko \cite{Sinyukov:2006dw,Gyulassy:2007zz,Sinyukov:qm08}. 
In this talk we point out the emergence of the {\em  initial azimuthally asymmetric} flow from FS+SE. 

\begin{figure}[tb]
\begin{center}
\subfigure{\includegraphics[angle=0,width= .9\textwidth]{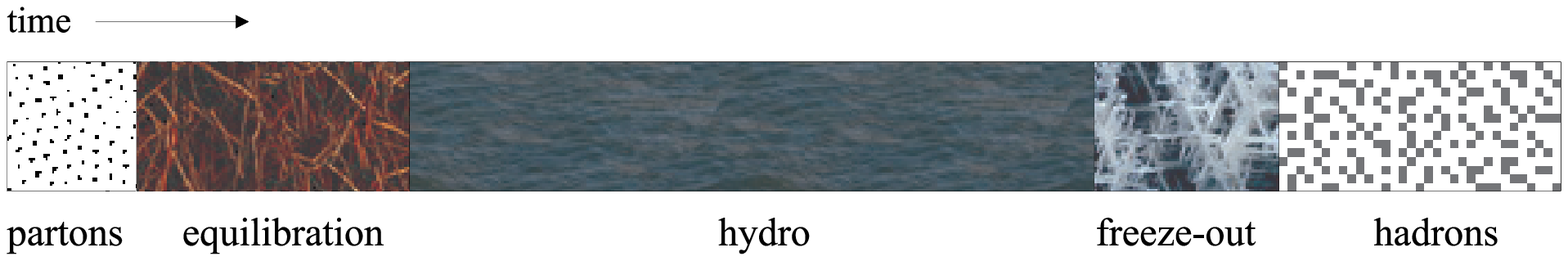}}\\
\vspace{2mm}
\subfigure{\includegraphics[angle=0,width= .9\textwidth]{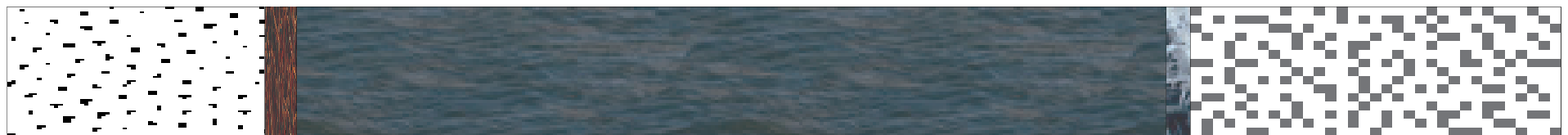}}
\end{center}
\caption{Evolution of the system formed in relativistic heavy-ion collisions, consisting of partonic free streaming, equilibration, hydrodynamics, freeze-out, and free streaming of hadrons to detectors. Top: equilibration and freeze-out occur gradually. Bottom: the approximation of
the sudden equilibration and instantaneous freeze-out. \label{fig:history}}
\end{figure}

The cartoon of the approach is given in Fig. \ref{fig:history}. 
Rather than assuming a gradual transition from an inequilibrated partonic stage to a thermalized system (top panel), in FS+SE one
approximates this early stage of evolution with free streaming followed by a sudden equilibration (bottom panel). Note that this is 
analogous to the standard treatment of the freeze-out at the end of the hydrodynamic evolution, where the continuous decoupling of hadrons 
(see the talks by Yu.~Sinyukov and J.~Knoll in these proceedings)
is approximated with a sudden Cooper-Frye freeze-out. 

Ever since the FS+SE has been proposed, it has been generally thought that it unavoidably 
reduces the elliptic flow, which develops hydrodynamically due to the azimuthal asymmetry of the initial density profile 
for non-central collisions. Admittedly, 
FS decreases the spatial azimuthal asymmetry of the system with time. We carefully reexamine this argument. 
The point is that while FS {\em alone} obviously cannot generate azimuthal asymmetry in the momentum distribution, 
due to the well-known fact that interactions among produced particles are needed to generate $v_2$, the SE
in fact does the job. This is because SE is dynamical in nature, resulting in 
an abrupt change, due to interactions, of the energy-momentum tensor of the system into a diagonal form 
(in the reference frame of the fluid element) of perfect hydro. 

One may interpret the FS+SE approach as an approximation to viscous hydrodynamics. 
Instead of considering a complicated viscous system far from the thermal equilibrium, where the partonic cross section  
has a finite value, one initially treats the partons as free, and 
later supplies them with a large cross section which results in an instantaneous equilibration of the system and transition to 
perfect hydrodynamics. Such an interpretation works when viscosity decreases with time, or equivalently, the 
partonic cross section increases. Confinement effects, which switch on as the distance between the partons increases, 
provide a mechanism for this behavior and support this interpretation, although a convincing solution of the 
early thermalization or isotropization problems is still missing despite many theoretical efforts.

\begin{figure}[tb]
\begin{center}
\subfigure{\includegraphics[angle=0,width=0.49 \textwidth]{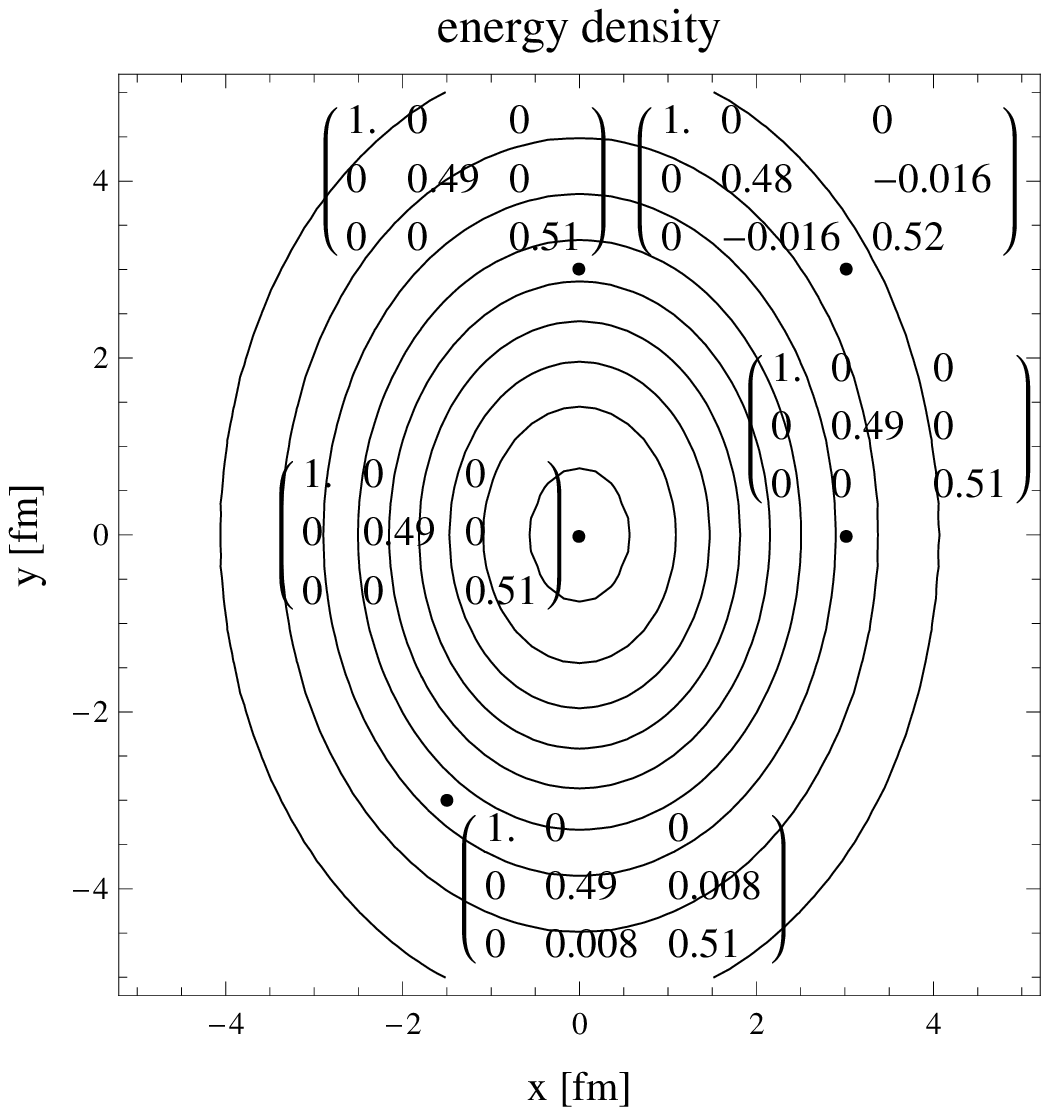}}
\subfigure{\includegraphics[angle=0,width=0.49 \textwidth]{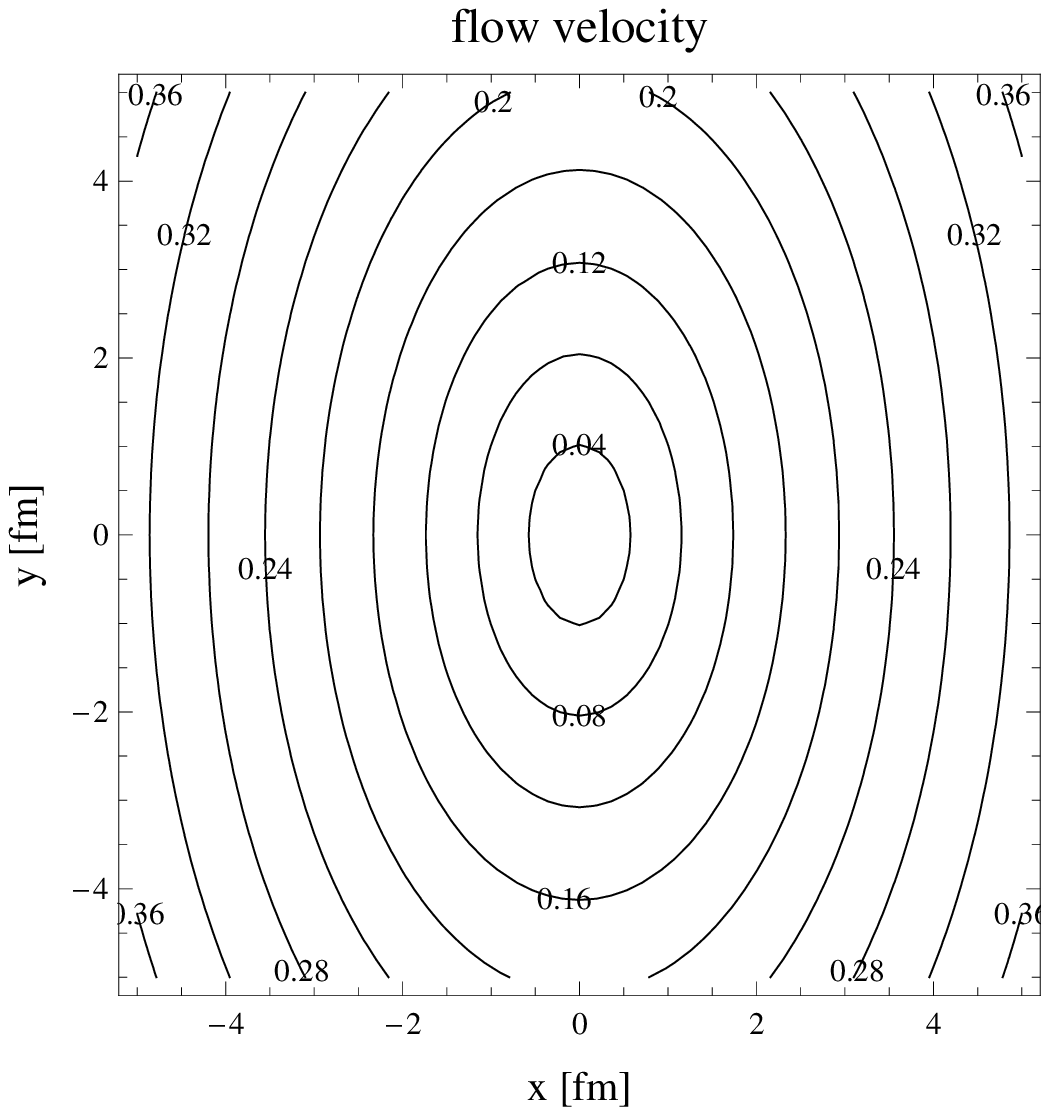}}
\end{center}
\vspace{-6.5mm}
\caption{
\label{fig:ploT} Left: Energy-density contours with the energy-momentum tensor in the local rest frame (in units of $\varepsilon$) shown 
at a few points (dots). Right: the profile of the transverse velocity, $v=\sqrt{v^2_x+v^2_y}$, in units of $c$.}
\end{figure}

Here we analyze a boost-invariant system with an initial Gaussian transverse energy profile, 
$n(x_0,y_0)=\exp \left ( -\frac{x_0^2}{2a^2} -\frac{y_0^2}{2 b^2} \right )$,
where the widths $a$ and $b$ depend on centrality and are obtained with {\tt GLISSANDO} \cite{Broniowski:2007nz}. 
We introduce the initial,
$\tau_0=\sqrt{t_0^2-z_0^2}$, and final, $\tau=\sqrt{t^2-z^2}$, 
proper times of free streaming, as well as the space-time rapidities 
\mbox{$\eta_0={1\over2}\log{{t_0-z_0}\over{t_0+z_0}}$} and \mbox{$\eta={1\over2}\log{{t-z}\over{t+z}}$}. Elementary kinematics, following 
from the fact that a massless parton moves along a straight line 
with the velocity of light and a four momentum \mbox{$p^\mu=(p_T {\rm cosh} Y, p_T \cos \phi, p_T \sin \phi, p_T {\rm sinh} Y)$}, 
relates the initial and final coordinates of the parton:
\begin{eqnarray}
&&\tau {\rm sinh}(\eta-Y)=\tau_0 {\rm sinh}(\eta_0-Y), \;\; x=x_0+d \cos \phi, \;\;  y=y_0 + d\sin \phi , \nonumber \\
&& d =\frac{t-t_0}{{\rm cosh}Y}=\tau {\rm cosh}(Y-\eta)-\sqrt{\tau_0^2+\tau^2 {\rm sinh}^2(Y-\eta)}.  \label{kinem} 
\end{eqnarray} 
Consequently, the phase-space densities of partons at the proper times $\tau_0$ and $\tau$ are related,
\begin{eqnarray}
&&\frac{d^6N(\tau)}{dY d^2p_T d\eta dx dy} = \int d \eta_0 dx_0 dy_0 \frac{d^6N(\tau_0)}{dY d^2p_T d\eta_0 dx_0 dy_0} \times \label{fs} \\ 
&& \delta(\eta_0-Y-{\rm arcsinh} [\frac{\tau}{\tau_0} {\rm sinh}(\eta-Y)] ) \delta(x - x_0 - d  \cos \phi)\delta( y - y_0 - d \sin \phi). \nonumber
\end{eqnarray}  
We assume for simplicity a factorized form of the initial parton distribution, 
\begin{eqnarray}
 \frac{d^6N(\tau_0)}{dY d^2p_T d\eta_0 dx_0 dy_0} = n(x_0,y_0) F(Y-\eta_0,p_T).
\end{eqnarray}
In Ref.~\cite{Broniowski:2008qk} we show that to a very good accuracy  
{$F(Y-\eta_0,p_T) \sim \delta(Y-\eta)$}. Then, the 
energy-momentum tensor of the system at $\eta=0$ is
\begin{eqnarray}
&&T^{\mu \nu}=A \int_0^{2 \pi} d\phi \, n\left(x-(\tau-\tau_0) \cos \phi,y-(\tau-\tau_0) \sin \phi\right) \times \nonumber \\
&&\left ( \begin{array}{cccc} 
1         &   \cos \phi         &  \sin \phi & 0\\ 
\cos \phi & \cos^2 \phi         & \cos \phi \sin \phi & 0\\ 
\sin \phi & \cos \phi \sin \phi & \sin^2 \phi & 0 \\
0 & 0 & 0 & 0 \end{array} \right ), \label{eq:T0}
\end{eqnarray}
where $A$ is a constant from the $p_T$ integration. 
Next, at each point we pass to the local reference frame in which the $T_{0i}$ components of the energy-momentum tensor vanish. 
The four-velocity needed for the appropriate boost is found from the condition  
\mbox{$T^{\mu \nu}(x,y) u_\nu(x,y) =\varepsilon(x,y) g^{\mu \nu} u_\nu(x,y)$}, 
where $\varepsilon$ is the energy density in the local rest frame.
In the left part of Fig.~\ref{fig:ploT} we show the profile of $\varepsilon$ together with 
the energy-momentum tensor in the local rest frame (in units of $\varepsilon$), displayed  
at a few points. We note that it has the structure very close to the case of the  
{\em perfect transverse hydrodynamics} of massless particles \cite{Bialas:2007gn}, where the transverse pressure is 
equal to $\varepsilon/2$. Small departures from this form, present in our case, have the structure of the shear 
tensor used to include the viscosity effects in transverse hydrodynamics 
\cite{Bozek:2007di}.

\begin{figure}[tb]
\begin{center}
\subfigure{\includegraphics[angle=0,width=0.44 \textwidth]{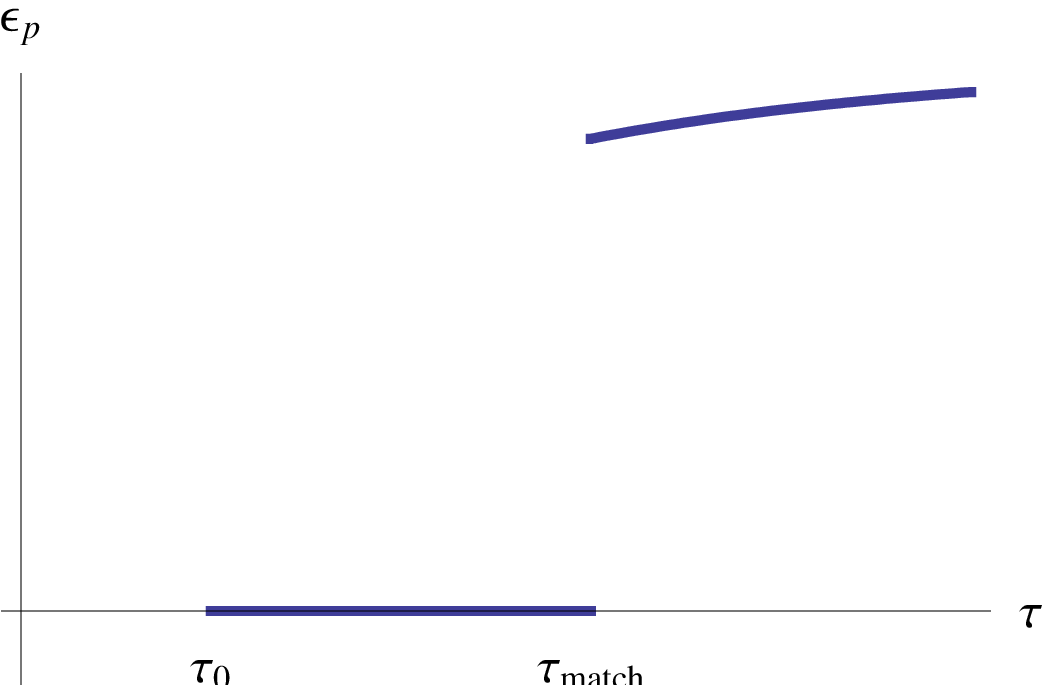}}
\subfigure{\includegraphics[angle=0,width=0.52 \textwidth]{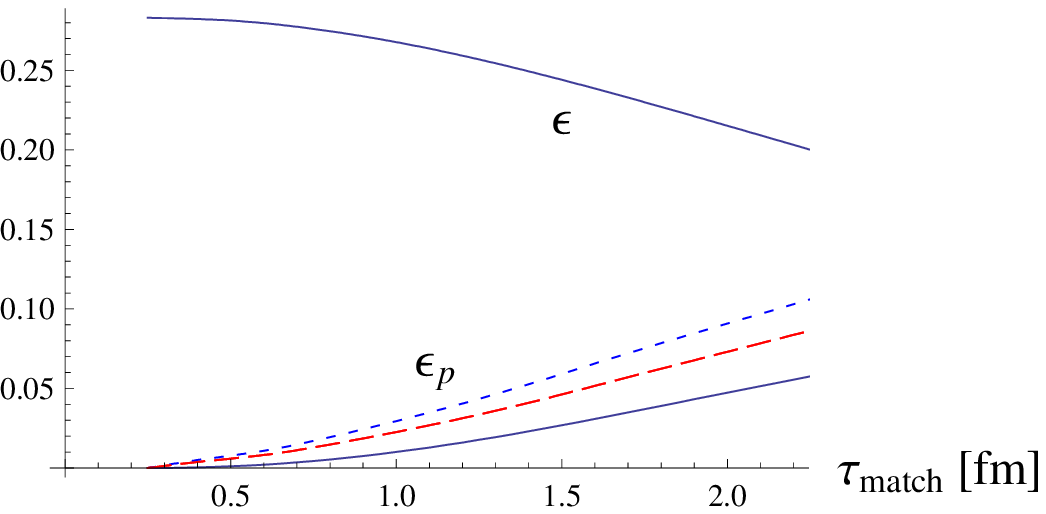}}
\end{center}
\vspace{-4.5mm}
\caption{Left: The schematic development of the partonic elliptic flow $\epsilon_p$ from FS+SE. 
Right: the value of the generated momentum asymmetry $\epsilon_p$ plotted as a function of the proper time when 
the Landau matching is imposed. 
The dotted (dashed) line corresponds to matching to isotropic (transverse) hydro, while the solid line shows 
the results of hydrodynamics only, with no FS present. 
The top curve shows the spatial asymmetry $\epsilon$, which decreases with time. 
\label{fig:match}} 
\end{figure}

The right part of Fig.~\ref{fig:ploT} shows the transverse velocity profile. 
We note that it is azimuthally asymmetric (stronger in the direction of the impact parameter), 
which simply reflects the original geometry. In fact, for low free-streaming times and close to the origin one finds 
\mbox{$ {\bf v}=-\frac{1}{3}(\tau-\tau_0) {\nabla n}/{n}$}. Thus the space -- collective velocity correlations are induced.

Following Ref.~\cite{Kolb:2000sd}, we now
consider a convenient measure of the {\em momentum} anisotropy,  
\mbox{$\epsilon_p=({\langle T_{xx}\rangle-\langle T_{yy}\rangle})/({\langle T_{xx}\rangle+\langle T_{yy}\rangle})$}, where brackets 
denote the spatial integration.
In the FS phase identically $\epsilon_p=0$, as no interactions have occurred. Then, at the proper time $\tau$, SE occurs. As a result, the 
energy momentum tensor in the local frame is replaced, due to interactions, with a diagonal form of the perfect hydrodynamics: 
$T^{\mu \nu} \to {\rm diag}(\varepsilon,  \varepsilon/3,  \varepsilon/3,  \varepsilon/3)$ for the isotropic, or  
$T^{\mu \nu} \to {\rm diag}(\varepsilon,  \varepsilon/2,  \varepsilon/2, 0)$ for the transverse hydro. 
It is this {\em Landau matching} condition which causes  $\epsilon_p$ to {\em jump to a nonzero value}. We show the result 
in Fig.~\ref{fig:match}.

\begin{figure}[tb]
\vspace{2mm}
\begin{center}
\includegraphics[angle=0,width=0.54 \textwidth]{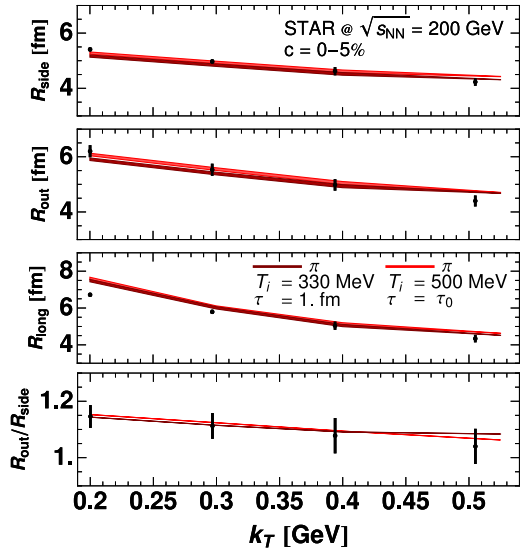}
\end{center}
\vspace{-1mm}
\caption{The pion HBT radii $R_{\rm side}$ , $R_{\rm out}$ , $R_{\rm long}$, 
and the ratio $R_{\rm out}/R_{\rm side}$ for central collisions. The darker (lighter) 
lines describe the results with (without) FS+SE. The data from Ref.~\cite{Adams:2004yc}.
\label{fig:hbt}}
\end{figure}

Finally, we compare the physical results obtained with (isotropic) hydrodynamics only (starting at an early proper 
time \mbox{$\tau_0=0.25$~fm}) to the results obtained with FS from $\tau_0$ to $\tau=1.0$~fm, followed by SE and 
(isotropic) hydro. We use the realistic equation of state \cite{Broniowski:2008vp}. 
The statistical hadronization is carried out with the help of {\tt THERMINATOR} \cite{Kisiel:2005hn}.
In Fig.~\ref{fig:hbt} we notice strikingly similar results for the two considered cases, 
not to mention the very good description of the HBT data, including the infamous ratio $R_{\rm out}/R_{\rm side}$. 
The similarity and agreement 
is similar for the $p_T$-spectra and $v_2$ \cite{Broniowski:2008qk,Broniowski:2008vp}, as well as for other 
centrality classes.
In Ref. \cite{Kisiel:2008ws} we have also shown 
that our model calculations reproduce very well the azHBT pion interferometry.

A practical conclusion from our study is that the inclusion of FS+SE may be used to {\em delay the start of perfect hydrodynamics} to 
``comfortable'' times of about 1~fm/c. The physical results remain
practically unaltered, since the decrease of the spatial anisotropy with time, 
resulting in milder hydrodynamic development of $v_2$, is intertwined with the buildup of the {\em initial 
azimuthally asymmetric} flow. Recall that in some studies, in order to obtain a proper description of the particle spectra and femtoscopy, 
hydro was used with initialization times of 0.1~fm/c \cite{Pratt:2008sz,Pratt:2008qv} (with viscous hydro). 
Recently, the phenomenological relevance of the initial flow, first examined in hydrodynamics in Ref.~\cite{Chojnacki:2004ec},
has been emphasized in Refs.~\cite{Pratt:2008qv,Lisa:2008gf} (see also the contribution of S.~Pratt to these proceedings).

\bigskip

Two of us (WB and WF) are grateful to Piotr Bo\.zek and Stanis\l{}aw Mr\'owczy\'nski for useful conversations.  


\begin{thebibliography}{10}

\bibitem{Kolb:2000sd}
P.F. Kolb, J. Sollfrank and U.W. Heinz,
\newblock Phys. Rev. C62 (2000) 054909, hep-ph/0006129.

\bibitem{Jas:2007rw}
W. Jas and S. Mrowczynski,
\newblock Phys. Rev. C76 (2007) 044905, 0706.2273.

\bibitem{Sinyukov:2006dw}
Y.M. Sinyukov,
\newblock Acta Phys. Polon. B37 (2006) 3343.

\bibitem{Gyulassy:2007zz}
M. Gyulassy et~al.,
\newblock Braz. J. Phys. 37 (2007) 1031.

\bibitem{Sinyukov:qm08}
Y. Sinyukov,
\newblock (2008),
\newblock {talk presented at Quark Matter 2008, Jaipur, India, 4-10 February
  2008}.

\bibitem{Broniowski:2007nz}
W. Broniowski, M. Rybczynski and P. Bozek,
\newblock Comput. Phys. Commun. 180 (2009) 69, 0710.5731.

\bibitem{Broniowski:2008qk}
W. Broniowski et~al.,
\newblock (2008), 0812.3393.

\bibitem{Bialas:2007gn}
A. Bialas, M. Chojnacki and W. Florkowski,
\newblock Phys. Lett. B661 (2008) 325, 0708.1076.

\bibitem{Bozek:2007di}
P. Bozek,
\newblock Acta Phys. Polon. B39 (2008) 1375, 0711.2889.

\bibitem{Adams:2004yc}
STAR, J. Adams et~al.,
\newblock Phys. Rev. C71 (2005) 044906, nucl-ex/0411036.

\bibitem{Broniowski:2008vp}
W. Broniowski et~al.,
\newblock Phys. Rev. Lett. 101 (2008) 022301, 0801.4361.

\bibitem{Kisiel:2005hn}
A. Kisiel et~al.,
\newblock Comput. Phys. Commun. 174 (2006) 669, nucl-th/0504047.

\bibitem{Kisiel:2008ws}
A. Kisiel et~al.,
\newblock (2008), 0808.3363.

\bibitem{Pratt:2008sz}
S. Pratt and J. Vredevoogd,
\newblock Phys. Rev. C78 (2008) 054906, 0809.0516.

\bibitem{Pratt:2008qv}
S. Pratt,
\newblock (2008), 0811.3363.

\bibitem{Chojnacki:2004ec}
M. Chojnacki, W. Florkowski and T. Csorgo,
\newblock Phys. Rev. C71 (2005) 044902, nucl-th/0410036.

\bibitem{Lisa:2008gf}
M.A. Lisa and S. Pratt,
\newblock (2008), 0811.1352.

\end{thebibliography}

\end{document}